\documentstyle[onecolumn,psfig]{mn}

\title{Correlation between radio and broad-line emissions in radio-loud quasars }
\author[X. Cao and D. R. Jiang]
       {Xinwu Cao and D. R. Jiang \\
Shanghai Observatory, Chinese Academy of Sciences, Shanghai,
200030, China; cxw@center.shao.ac.cn}
           
\date{Accepted 1999 March 18. Received 1999 February 8; in original form
1999 January 5 }

\pagerange{\pageref{firstpage}--\pageref{lastpage}}
\pubyear{}

\begin{document}
\maketitle
\label{firstpage}

\begin{abstract}
We present a correlation between radio and broad-line emission for a sample
of radio-loud quasars, which is in favour of a close link between the
accretion processes and the relativistic jets. The BL Lac objects seem
to follow the statistical behaviour of the quasars, but with
fainter broad-line emission. 

\end{abstract}


\section{Introduction}

The relation between the jets and the accretion processes in the central
'engine' is a crucial ingredient in our understanding of the physics
of active galactic nuclei (AGN). In some theoretical models
of the formation of the jet, the power is generated
through accretion and then extracted from the disc/black hole rotational
energy and converted into the kinetic power of the jet (Blandford \& Znajek; 
Blandford \& Payne 1982). Recently, 
the concept of jet-disc symbiosis was introduced and the
inhomogeneous jet model plus mass and energy conservation in the jet-disc 
system was applied to study the relation between disc and
jet luminosities (Falcke \& Biermann 1995; Falcke, Malkan \& Biermann 1995;
Falcke \& Biermann 1998). 
An effective  approach to study the link between
these two phenomena is to explore the relationship between luminosity
in line emission and kinetic power of jets in different
scales (Rawlings \& Saunders 1991; Celotti, Padovani \& Ghisellini 1997; Cao
\& Jiang 1998). Rawlings \& Saunders (1991) derived the total jet kinetic
power $Q_{jet}$ and found a correlation between $Q_{jet}$ and the narrow
line luminosity $L_{NLR}$. There are some indications that the narrow-line
emission could be partly caused by the power supplied by the jet (Lacy \&
Rawlings 1994; Capetti et al. 1996). Celotti, Padovani \& Ghisellini (1997)
considered the luminosity in broad emission lines and explored its relation
with kinetic power for a sample of radio-loud objects.
They also found  evidence for a link between jets and discs.
The kinetic power of jets can be estimated by 
using radio data on very long-baseline interferometry (VLBI) scales and the
standard synchrotron self-Compton theory
(Celotti \& Fabian 1993), or in a similar way based on K\"onigl's
inhomogeneous jet model (Cao \& Jiang 1998).
The kinetic power of the jet can be estimated in this way only for a small
fraction of quasars due to the lack of the necessary observational data for
many sources.

Serjeant et al. (1998) found a significant correlation between radio and
optical emission for a sample of steep-spectrum quasars,
which  presents  direct evidence for a close
link between accretion on to  black holes and the fuelling of relativistic
jets. Their sample is limited
to steep-spectrum quasars to reduce the effects of relativistic beaming for
optical continuum.

The luminosity in the 
broad-line region  can be taken as an indicator of accretion power of the
source (Celotti, Padovani \& Ghisellini 1997), and the radio luminosity
is believed to be a straightforward indicator of jet power (Serjeant et al.
1998). In this paper we present a correlation between radio and broad-line
emission for a sample of radio-loud quasars.
The sample and the estimate of total broad-line flux are
described in Sect. 2. Section 3 contains the results. The last section
is devoted to discussion. The cosmological parameters $H_{0}=50$ km
s$^{-1}$ Mpc$^{-1}$ and $q_{0}=$0.5 have been adopted in this work.

\section{The  sample}

The 1 Jy catalogue is an all-sky survey covering 9.81 sr with a flux density
limit of $S_{\rm 5 GHz}\ge 1$ Jy lying outside the galactic plane ($|b|\ge
10^{\circ}$) (K\"uhr et al. 1981b).
Optical counterparts have been found for 97 \% of the radio sources in
1 Jy catalogue (Stickel, Meisenheimer \& K\"uhr 1994). 
The S4 survey covers the region between $35^{\circ} \le \delta \le 70^{\circ}$
with $S_{\rm 5 GHz}\ge 0.5$ Jy (Pauliny-Toth et al. 1978), 
while the S5 survey contains the sources with $S_{\rm 5 GHz}\ge 0.25$
Jy in the region $\delta \ge 70^{\circ}$ (K\"uhr et al. 1981a). 
For the S4 catalogue, about 90 \% of the radio sources have known optical
counterparts (Stickel \& K\"uhr 1994), while about 75 \% of the sources
have optical counterparts in S5 catalogue (Stickel \& K\"uhr 1996a).

In this work, we only consider the quasars and BL Lac objects
in the 1 Jy, S4 and S5 radio source catalogues (all sources
identified as galaxies have not been considered here). 
Combining these three catalogues, we find 378 sources
with available redshifts including 358 quasars and 20 BL Lac objects
(only those identified as BL/QSO).
Complete information on the line spectra is available for very few 
sources in our sample, since different lines are observed for the sources
at different redshifts.
We have to estimate the total broad-line flux from the available observational
data. There is not a solidly established procedure to derive the total
broad-line flux and we therefore adopt the method proposed by Celotti,
Padovani \& Ghisellini (1997).
The following lines: Ly$\alpha$, C\,{\sc iv},  Mg\,{\sc ii}, H$\gamma$,
H$\beta$, and H$\alpha$, which contribute the major parts in the total
broad-line emissions, are used in our estimate. We use the line ratios reported by Francis et al.
(1991) and add the contribution from line H$\alpha$ to derive the total
broad-line flux (see Celotti, Padovani \& Ghisellini 1997 for details).
We then search the literature to collect data on broad-line fluxes.
We only consider values of line fluxes (or luminosities) given directly
or the equivalent width and the continuum flux at the corresponding line
frequency which are reported together in the literatures (slightly
different from Celotti, Padovani \& Ghisellini (1997)). When more than one
value of the same line flux was found in the literature, we take the most
recent reference.
This finally leads to a sample  of 198
sources, consisting of 184 quasars and 14 BL Lac objects.
Our sample covers more than 50 \% of the 378 sources in the starting
samples. The radio data in the sample are taken from some related references
(Stickel \& K\"uhr 1996a, 1996b; Stickel \& K\"uhr 1994; Stickel,
Meisenheimer \& K\"uhr 1994). The radio data and broad-line fluxes 
are listed in Table 1.

\begin{figure}
\centerline{\psfig{figure=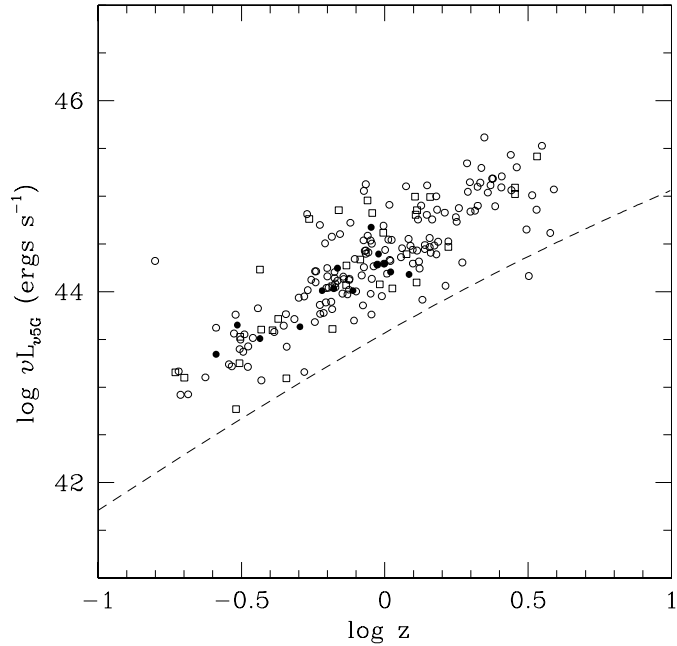,width=10.0cm,height=10.0cm}}
\caption{The radio luminosity $\nu L_{\nu{5G}}$, redshift $z$ plane for
the sample. The dashed line denotes the low flux density limit.
The open circles represent the quasars with $\alpha_{11-6}> -0.7$, and the
squares represent the quasars with $\alpha_{11-6}\le -0.7$, while the full circles
represent BL Lac objects. }
\end{figure}

\begin{figure}
\centerline{\psfig{figure=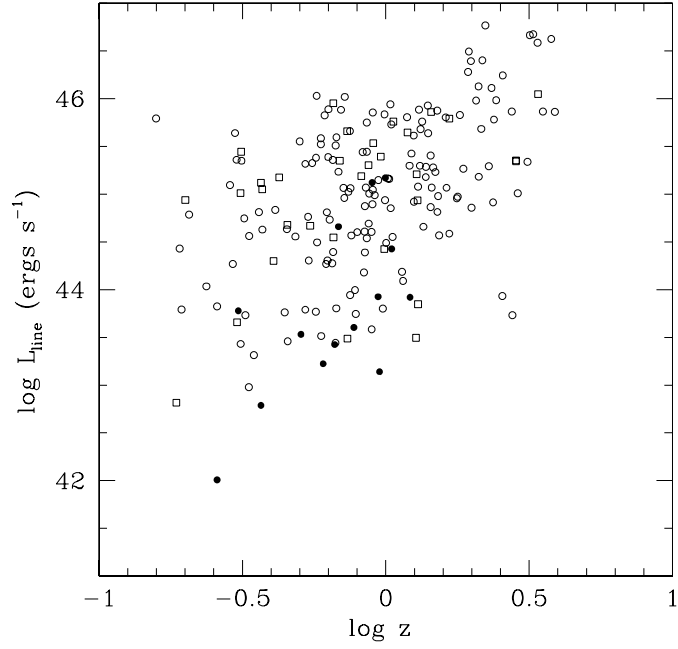,width=10.0cm,height=10.0cm}}
\caption{The total broad-line luminosity $L_{\rm line}$, redshift $z$ plane
for the sample. (Symbols as in Fig. 1. }
\end{figure}

\begin{figure}
\centerline{\psfig{figure=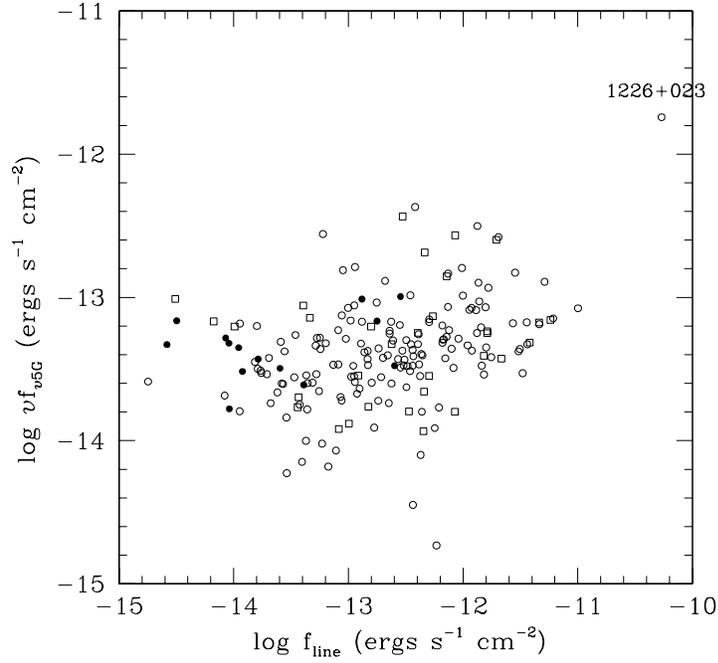,width=10.0cm,height=10.0cm}}
\caption{The radio  and broad-line flux relation (symbols as in Fig. 1.).}
\end{figure}

\begin{figure}
\centerline{\psfig{figure=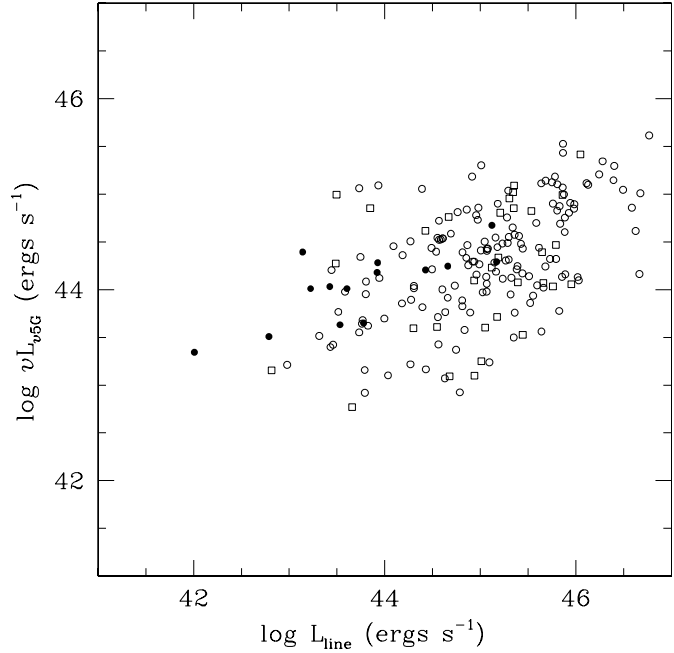,width=10.0cm,height=10.0cm}}
\caption{The radio  and broad-line luminosity relation (symbols as in Fig. 1.).}
\end{figure}

\newpage
\begin{table}
 \begin{minipage}{150mm}
  \caption{Radio and BLR data of the sample.}
  \begin{tabular}{ccccllrr}\hline
Source & Class. & z & log f$_{BLR}$ & Lines & Refs. & f$_{5{\rm G}}$ & $\alpha_{11-6}$ \\
(1) & (2) & (3) & (4) & (5) &(6) & (7) & (8)\\ \hline
 0003$-$066& Q &0.347 & -13.46& H$\alpha$ & S89& 1.48 & 0.02\\ 
 0014+813& Q &3.384 & -12.37& Ly$\alpha$, C\,{\sc iv} &O94 & 0.55 & -0.16\\ 
 0016+731& Q &1.781 & -13.36& Ly$\alpha$, C\,{\sc iv}, Mg\,{\sc ii}& L96& 1.65 & 0.16\\ 
 0022$-$297& Q &0.406 & -12.62& Mg\,{\sc ii}, H$\beta$, H$\alpha$ & S93b& 1.03 & -0.74\\ 
 0024+348& Q &0.333 & -13.76& H$\beta$, H$\alpha$ & SK93b & 0.73 & -0.30\\ 
 0035+413& Q &1.353 & -13.40& Mg\,{\sc ii} & SK93b & 0.64 & 0.75\\ 
 0056$-$001& Q &0.717 & -12.50& Mg\,{\sc ii}& B89& 1.40 & -0.39\\ 
 0106+013& Q &2.107 & -12.36& Ly$\alpha$, C\,{\sc iv}& B89& 2.28 & -0.08\\ 
 0112$-$017& Q &1.381 & -12.80& C\,{\sc iv}, Mg\,{\sc ii}&B89 & 1.20 & -0.01\\ 
 0119+041& Q &0.637 & -12.61& Mg\,{\sc ii}, H$\gamma$, H$\beta$&JB91,RS80 & 1.67 & 0.04\\ 
 0119+115& Q &0.570 & -13.47& Mg\,{\sc ii} &S89 & 1.01 & 0.33\\ 
 0133+207& Q &0.425 & -11.79& Ly$\alpha$, C\,{\sc iv}, H$\gamma$, H$\beta$ &W95,JB91 & 1.08 &-1.11\\ 
 0133+476& Q &0.859 & -13.09& Mg\,{\sc ii}, H$\gamma$, H$\beta$ &L96 & 3.22 & 0.62\\ 
 0134+329& Q &0.367 & -11.71& Ly$\alpha$, H$\gamma$, H$\beta$, H$\alpha$ &JB91,K85 & 5.28 & -0.86\\ 
 0135$-$247& Q &0.831 & -12.16& H$\gamma$, H$\beta$ & JB91& 1.70 & 0.35\\ 
 0153+744& Q &2.338 & -12.48& Ly$\alpha$, C\,{\sc iv}, Mg\,{\sc ii}& L96& 1.52 & -0.31\\ 
 0159$-$117& Q &0.669 & -11.88& Mg\,{\sc ii}, H$\beta$ & O84& 1.39 & -0.59\\ 
 0210+860& Q &0.186 & -13.39& H$\gamma$, H$\beta$, H$\alpha$ & L96& 1.72 &-1.13\\ 
 0212+735& Q &2.367 & -13.69& Ly$\alpha$, C\,{\sc iv}, Mg\,{\sc ii}&L96 & 2.20 & -0.12\\ 
 0229+131& Q &2.065 & -12.49& Ly$\alpha$, C\,{\sc iv} &O94 & 1.00 & -0.33\\ 
 0234+285& Q &1.210 & -12.66& C\,{\sc iv} & W84& 1.44 & -0.24\\ 
 0235+164& BL& 0.940 &  -13.79& Mg\,{\sc ii} & S93a  & 2.85 & 1.03\\ 
 0237$-$233& Q &2.224 & -11.78& H$\beta$, H$\alpha$ & B94& 3.40 & -0.68\\
 0248+430& Q &1.316 & -12.74& Mg\,{\sc ii} & S93b& 1.20 & 0.37\\ 
 0256+075& Q &0.893 & -14.08& Mg\,{\sc ii} & S89& 1.04 & 0.45\\ 
 0336$-$019& Q &0.852 & -12.55 & Mg\,{\sc ii}, H$\gamma$, H$\beta$ &B89,JB91 & 2.86 & 0.30\\ 
 0400+258& Q &2.109 & -13.31& C\,{\sc iv} &W84 & 1.79 & 0.11\\ 
 0403$-$132& Q &0.571 & -11.86& C\,{\sc iv}, Mg\,{\sc ii}, H$\gamma$, H$\beta$& JB91,M96,O84& 2.88 & -0.05\\ 
 0405$-$123& Q &0.574 & -11.22& Ly$\alpha$, C\,{\sc iv}, Mg\,{\sc ii}, H$\beta$, H$\alpha$ &O94,O84,M96,N79 & 1.96 & -0.30\\ 
 0414$-$189& Q &1.536 & -13.62& C\,{\sc iv}, Mg\,{\sc ii} &H78 & 1.35 & 0.22\\ 
 0420$-$014& Q &0.915 & -12.70& Mg\,{\sc ii} &B89 & 1.46 & 0.01\\ 
 0437+785& Q &0.454 & -12.35& H$\gamma$, H$\beta$ & S93b & 0.26 & -0.72\\ 
 0440$-$003& Q &0.844 & -13.00& Mg\,{\sc ii}, H$\gamma$, H$\beta$ &B89,JB91 & 2.61 & -0.29\\ 
 0440+732& Q &1.290 & -13.08& Mg\,{\sc ii} & S93b & 0.30 & -0.75\\ 
 0444+634& Q &0.781 & -13.54& Mg\,{\sc ii}, H$\beta$ & SK93b & 0.50 & -0.04\\ 
 0451$-$282& Q &2.559 & -12.44& Ly$\alpha$, C\,{\sc iv} & F83& 2.26 & -0.04\\ 
 0454$-$810& Q &0.444 & -13.24& Mg\,{\sc ii}, H$\beta$, H$\alpha$ & S89& 1.40 & 0.29\\ 
 0454$-$463& Q &0.858 & -12.18& Mg\,{\sc ii} & F83 & 2.32 & -0.03 \\
 0454$-$234& Q &1.003 & -13.29& Mg\,{\sc ii} & S89& 2.06 & 0.16\\ 
 0457+024& Q &2.384 & -12.83& Ly$\alpha$, C\,{\sc iv} & B89& 1.15 & -0.64\\ 
 0458$-$020& Q &2.286 & -13.28& Ly$\alpha$, C\,{\sc iv} & B89& 1.74 & -0.08\\ 
 0511$-$220& Q &1.296 & -12.94& C\,{\sc iv}, Mg\,{\sc ii} &S89 & 1.31 & 0.13\\ 
 0514$-$459& Q &0.194 & -12.45& H$\beta$, H$\alpha$ & S93b& 1.06 & -0.32\\ 
 0518+165& Q &0.759 & -12.94& H$\gamma$, H$\beta$ & JB91& 4.11 & -0.59\\
 0528$-$250 & Q& 2.765 & -15.03 & Ly$\alpha$ & WM96 & 1.16 & -0.20 \\  
 0537$-$441 &BL & 0.896 &  -12.55& Mg\,{\sc ii}  & S93a  & 4.00 & 0.06\\ 
 0537$-$286 & Q& 3.119 &  -13.54 & Ly$\alpha$, C\,{\sc iv} & O94 & 1.02 & 0.52\\ 
 0538+498& Q &0.545 & -12.53& Mg\,{\sc ii}, H$\gamma$, H$\beta$ &L96 & 8.30 & -0.72\\
 0539$-$057& Q &0.839 & -13.42& Mg\,{\sc ii} & SK93b& 1.55 & 1.41\\ 
 0602$-$319& Q &0.452 & -12.39& H$\beta$ &R84 & 1.25 & -0.67\\ 
 0605$-$085& Q &0.872 & -12.95& Mg\,{\sc ii} & S93b& 3.49 & 0.09\\ 
 0607$-$157& Q &0.324 & -12.98& H$\beta$, H$\alpha$ &H78 & 1.82 & -0.01\\ 
 0636+680& Q &3.180 & -12.23& Ly$\alpha$, C\,{\sc iv} & O94 & 0.54 & 0.87\\ 
 0637$-$752& Q &0.654 & -12.01& Ly$\alpha$, C\,{\sc iv}, H$\gamma$, H$\beta$ & W98,H78,T93& 5.85 & 0.19\\ 
 0642+449& Q &3.396 & -12.91& Ly$\alpha$, C\,{\sc iv} & O94 & 0.78 & -0.79\\ 
 0646+600& Q &0.455 & -13.57& H$\gamma$, H$\beta$ & SK93b & 0.79 & 0.23\\ 
 0707+689& Q &1.139 & -13.71& Mg\,{\sc ii} & SK93b & 0.75 & -0.67\\ 
 0711+356& Q &1.620 & -12.43& C\,{\sc iv}, Mg\,{\sc ii} &L96 & 1.52 & -0.29\\ 
 0723+679& Q &0.846 & -12.74& Mg\,{\sc ii}, H$\gamma$, H$\beta$ &L96 & 1.32 & -0.34\\ 
 0736+017& Q &0.191 & -11.80& Ly$\alpha$, C\,{\sc iv}, H$\gamma$, H$\beta$, H$\alpha$ & JB91,O94& 1.99 & -0.13\\ 
 0738+313& Q &0.631 & -11.45& Mg\,{\sc ii}, H$\gamma$, H$\beta$ &W85,JB91 & 2.49 & 0.27\\ 
 \end{tabular}
\end{minipage}
\end{table}

\begin{table}
 \begin{minipage}{150mm}
  \contcaption{Radio and BLR data of the sample.}
  \begin{tabular}{ccccllrr}\hline
 0740+380& Q &1.063 & -12.07& Ly$\alpha$, C\,{\sc iv} & W95 & 0.27 &-1.24\\ 
 0740+828& Q &1.041 & -12.96& Mg\,{\sc ii} & S93b & 0.93 & -0.53\\ 
 0743$-$673& Q &1.511 & -12.30& C\,{\sc iv}, Mg\,{\sc ii}, H$\alpha$ & E89,d94 & 1.79 & -0.69\\ 
 0743$-$006& Q &0.994 & -12.83& Mg\,{\sc ii} & S89& 1.99 & 0.57\\ 
 0743+744& Q &1.629 & -13.18& Mg\,{\sc ii} & S93b & 0.33 & -0.05\\ 
 0804+499& Q &1.433 & -12.71& C\,{\sc iv}, Mg\,{\sc ii} &L96 & 2.05 & 0.47\\ 
 0809+483& Q &0.871 & -12.34& Mg\,{\sc ii}, H$\gamma$, H$\beta$ & L96& 4.42 & -0.89\\ 
 0814+425& BL &0.258 & -14.50& Mg\,{\sc ii} & L96& 1.69 & -0.10\\
 0820+225& BL &0.951 & -14.58& Mg\,{\sc ii} & S93a& 1.60 & -0.20\\ 
 0823+033& BL &0.506 & -13.60& Mg\,{\sc ii} & S93a& 1.32 & 0.77\\ 
 0825$-$202& Q &0.822 & -12.40& Mg\,{\sc ii} & S93b& 1.18 & -0.93\\ 
 0834$-$201& Q &2.752 & -12.89& Ly$\alpha$, C\,{\sc iv} & F83& 3.72 & 0.03\\ 
 0836+710& Q &2.172 & -12.12& Ly$\alpha$, C\,{\sc iv}, Mg\,{\sc ii} & L96& 2.59 & -0.32\\ 
 0838+133& Q &0.684 & -12.18& Ly$\alpha$, C\,{\sc iv}, H$\gamma$, H$\beta$ &W95,JB91,M96 & 1.39 & -0.39\\ 
 0842$-$754& Q &0.524 & -11.84& H$\beta$ & T93& 1.42 & -0.67\\ 
 0850+581& Q &1.322 & -12.36& Mg\,{\sc ii} &L96 & 1.39 & 0.75\\ 
 0851+202& BL &0.306 & -12.88& Mg\,{\sc ii}, H$\beta$, H$\alpha$& S89,S93a& 2.62 & 0.11\\ 
 0858$-$279& Q &2.152 & -12.83 & Ly$\alpha$, C\,{\sc iv} & d94& 1.42 & -0.55\\ 
 0859$-$140& Q &1.339 & -12.29& H$\alpha$ & E89& 2.30 & -0.42\\ 
 0859+470& Q &1.462 & -12.86& C\,{\sc iv}, Mg\,{\sc ii} &L96 & 1.78 & -0.15\\
 0906+015& Q &1.018 & -12.62& Mg\,{\sc ii} & B89& 1.04 & 0.04\\ 
 0906+430& Q &0.668 & -13.95& Mg\,{\sc ii}, H$\gamma$, H$\beta$ &L96 & 1.80 & -0.38\\ 
 0917+624& Q &1.446 & -13.06& Mg\,{\sc ii} & SK93b& 1.00 & 0.08\\ 
 0923+392& Q &0.698 & -11.55& Ly$\alpha$, C\,{\sc iv}, Mg\,{\sc ii}, H$\gamma$, H$\beta$ & W95,L96& 8.73 & 1.03\\ 
 0945+408& Q &1.252 & -12.37& C\,{\sc iv}, Mg\,{\sc ii} &L96 & 1.39 & 0.11\\ 
 0953+254& Q &0.712 & -12.38& H$\gamma$, H$\beta$ & JB91& 1.82 & 0.83\\ 
 0954+556& Q &0.901 & -12.63& Ly$\alpha$, C\,{\sc iv}, Mg\,{\sc ii}, H$\gamma$, H$\beta$ & W95,L96& 2.28 & -0.19\\ 
 0954+658& BL &0.367 & -14.04& H$\alpha$ & L96& 1.46 & 0.35\\ 
 1007+417& Q &0.6123 & -11.48& C\,{\sc iv}, H$\gamma$, H$\beta$ & JB91,M96& 0.71 & -0.63\\ 
 1007+716& Q &1.192 & -12.30& Mg\,{\sc ii} & S93b & 0.59 & -0.95\\ 
 1017$-$426& Q &1.280 & -12.80& Mg\,{\sc ii} & S93b& 1.27 & -0.98\\ 
 1034$-$293& Q &0.312 & -13.25& H$\beta$, H$\alpha$ &S89 & 1.51 & 0.35\\ 
 1040+123& Q &1.029 & -12.64& Mg\,{\sc ii} & N79& 1.47 & -0.61\\ 
 1045$-$188& Q &0.595 & -13.77& H$\beta$ & S93b& 1.14 & 0.32\\ 
 1055+018& Q &0.892 & -13.06& Mg\,{\sc ii} & B89& 3.47 & 0.32\\ 
 1100+772& Q &0.3115 & -11.67& Ly$\alpha$, C\,{\sc iv}, H$\gamma$, H$\beta$, H$\alpha$ & JB91,W98,M96& 0.77 & -0.88\\ 
 1111+408& Q &0.734 & -11.82& Ly$\alpha$, C\,{\sc iv}, H$\gamma$, H$\beta$ & JB91,W95 & 0.79 & -0.98\\ 
 1127$-$145 & Q& 1.187 &  -12.13& Ly$\alpha$ & W95  & 6.57 & 0.03\\ 
 1136$-$135& Q &0.554 & -11.89& Mg\,{\sc ii}, H$\beta$ &O84,T93 & 2.11 & -0.42\\
 1137+660& Q &0.6563 & -11.42& Ly$\alpha$, C\,{\sc iv}, Mg\,{\sc ii}, H$\gamma$, H$\beta$ & W95,M96,JB91& 1.06 & -0.82\\ 
 1144$-$379& BL &1.048 & -13.39& Mg\,{\sc ii} &S89 & 1.61 & 0.66\\ 
 1148$-$001& Q &1.982 & -12.04& Ly$\alpha$, C\,{\sc iv} &B89 & 1.90 & -0.44\\ 
 1150+497& Q &0.334 & -12.18& H$\beta$ & SM87& 1.12 & -0.48\\ 
 1151$-$348& Q &0.258 & -12.68& H$\beta$ &T93 & 2.83 & -0.65\\ 
 1226+023& Q &0.158 & -10.27& Ly$\alpha$, C\,{\sc iv}, H$\beta$, H$\alpha$ &O94,JB91,M96 &42.85 & 0.15\\ 
 1229$-$021& Q &1.045 & -12.08& Mg\,{\sc ii} & B89& 1.07 & -0.29\\ 
 1237$-$101& Q &0.753 & -12.44& Mg\,{\sc ii} &S93b & 1.31 & -0.25\\ 
 1245$-$197 &Q&1.275 &-14.51 &Mg\,{\sc ii} &d94 & 2.50 &-0.70 \\
 1250+568& Q &0.321 & -11.96& C\,{\sc iv}, Mg\,{\sc ii}, H$\gamma$, H$\beta$, H$\alpha$ & JB91,W95& 1.06 & -0.51\\ 
 1253$-$055& Q &0.536 & -12.42& Ly$\alpha$, C\,{\sc iv}, Mg\,{\sc ii}, H$\beta$, H$\alpha$ & W95,M96,N79&14.95 & 0.30\\ 
 1258+404& Q &1.6656 & -12.47& C\,{\sc iv}, Mg\,{\sc ii} & C91 & 0.36 & -0.88\\ 
 1302$-$102& Q &0.286 & -11.51& Ly$\alpha$, C\,{\sc iv}, H$\beta$ &O94,M96 & 1.17 & 0.17\\ 
 1308+326& BL &0.997 & -12.60& Ly$\alpha$, Mg\,{\sc ii} & S93a,O94& 1.53 & 0.20\\ 
 1328+307& Q &0.846 & -13.22& H$\gamma$, H$\beta$ &GW94 & 7.40 & -0.53\\ 
 1334$-$127& Q &0.539 & -12.88& Mg\,{\sc ii} & S93b& 2.24 & 0.17\\ 
 1340+606& Q &0.961 & -12.34& Ly$\alpha$, C\,{\sc iv} & W95 & 0.40 &-1.14\\ 
 1354+195& Q &0.720 & -11.44& Ly$\alpha$, C\,{\sc iv} & W95&1.56 & -0.07\\ 
 1355$-$416& Q &0.313 & -11.24& Ly$\alpha$, H$\beta$ &T93,O94 & 1.44 & -0.88\\ 
 1416$-$067& Q &1.439 & -12.26& H$\alpha$ & E89& 1.53 & -0.96\\ 
 1424$-$418& Q &1.522 & -13.20& Mg\,{\sc ii} & S89& 3.13 & 0.28\\ 
 1442+101& Q &3.5305 & -13.13& C\,{\sc iv} & C91& 1.22 & -0.61\\ 
 1451$-$375& Q &0.314 & -11.34& Ly$\alpha$ & K85& 1.89 & 0.37\\ 
 1452+502& Q &2.849 & -13.43& Ly$\alpha$, C\,{\sc iv} & SK93a & 0.54 & -0.78\\ 
 1458+718& Q &0.905 & -12.14& Mg\,{\sc ii}, H$\gamma$, H$\beta$ & L96& 3.39 & -0.71\\ 
 1504$-$166& Q &0.876 & -12.64& Mg\,{\sc ii} & H78& 1.98 & -0.16\\

 \end{tabular}
\end{minipage}
\end{table}

\begin{table}
 \begin{minipage}{150mm}
  \contcaption{Radio and BLR data of the sample.}
  \begin{tabular}{ccccllrr}\hline
 1510$-$089& Q &0.361 & -12.00& Ly$\alpha$, Mg\,{\sc ii}, H$\gamma$, H$\beta$, H$\alpha$ & N79,T93,O94,BK84 & 3.08 & 0.31\\ 
 1512+370& Q &0.371 & -12.21& Ly$\alpha$, C\,{\sc iv}, H$\beta$ & M96,W98& 0.41 & -0.45\\ 
 1518+046& Q &1.296 & -14.17& Mg\,{\sc ii} & d94 & 1.06 & -1.30\\
 1531+722& Q &0.899 & -12.78& Mg\,{\sc ii} & S93b & 0.46 & -0.04\\ 
 1532+016& Q &1.435 & -13.26& C\,{\sc iv}, Mg\,{\sc ii} &B89 & 1.14 & 0.06\\ 
 1538+149& BL &0.605 & -14.07& Mg\,{\sc ii} & S93a& 1.96 & 0.34\\ 
 1546+027& Q &0.412 & -12.10& Mg\,{\sc ii} & B89& 1.45 & 0.46\\ 
 1555+001& Q &1.770 & -13.37& Ly$\alpha$, C\,{\sc iv} & B81 & 2.24 & 0.34\\ 
 1611+343& Q &1.401 & -12.17& Ly$\alpha$, C\,{\sc iv} & W95& 2.67 & 0.10\\
 1622$-$253 &Q& 0.786 & -13.80 & Mg\,{\sc ii}, H$\beta$ &d94 &2.08 &-0.14 \\
 1624+416& Q &2.550 & -14.75& Ly$\alpha$, C\,{\sc iv} & L96& 1.32 & -0.26\\ 
 1633+382& Q &1.814 & -12.52& Ly$\alpha$, C\,{\sc iv}, Mg\,{\sc ii} &L96 & 4.02 & 0.73\\ 
 1634+628& Q &0.988 & -13.34& Mg\,{\sc ii}, H$\gamma$, H$\beta$ &L96 & 1.52 & -0.92\\ 
 1637+574& Q &0.750 & -11.84& Ly$\alpha$, C\,{\sc iv}, Mg\,{\sc ii}, H$\gamma$, H$\beta$ & W95,L96& 1.42 & 0.35\\ 
 1638+398& Q &1.666 & -13.68& C\,{\sc iv}, Mg\,{\sc ii} &S89 & 1.15 & 0.17\\ 
 1641+399& Q &0.594 & -11.69& Ly$\alpha$, C\,{\sc iv}, Mg\,{\sc ii}, H$\gamma$, H$\beta$ & L96,W95&10.81 & 0.54\\ 
 1642+690& Q &0.751 & -13.56& Mg\,{\sc ii}, H$\gamma$, H$\beta$ &L96 & 1.39 & -0.10\\ 
 1704+608& Q &0.371 & -11.79& Ly$\alpha$, C\,{\sc iv}, H$\beta$, H$\alpha$ &N79,W98,M96 & 1.23 & -0.81\\ 
 1721+343& Q &0.206 & -11.52& Ly$\alpha$, C\,{\sc iv}, H$\gamma$, H$\beta$, H$\alpha$ & W98,R84,S81 & 0.93 & -0.43\\ 
 1725+044& Q &0.293 & -12.35& H$\beta$ & R84& 1.24 & 0.76\\
 1732+389& Q &0.976 & -13.95& Mg\,{\sc ii} & S89& 1.13 & 0.85\\ 
 1739+522& Q &1.379 & -12.90& C\,{\sc iv}, Mg\,{\sc ii} & L96& 1.98 & 0.68\\ 
 1741$-$038& Q &1.057 & -13.27& Mg\,{\sc ii} & S89& 3.68 & 0.75\\ 
 1745+624& Q &3.886 & -13.23& Ly$\alpha$, C\,{\sc iv} & SK93a & 0.58 & -0.30\\ 
 1803+784& BL &0.684 & -12.75& Mg\,{\sc ii}, H$\beta$ & L96& 2.62 & 0.25\\ 
 1823+568& BL &0.664 & -13.96& Mg\,{\sc ii}, H$\beta$  & L96& 1.66 & 0.22\\ 
 1828+487& Q &0.691 & -12.07& Mg\,{\sc ii}, H$\gamma$, H$\beta$ &L96 & 6.19 & -0.75\\ 
 1830+285& Q &0.594 & -11.76& Mg\,{\sc ii} & RS80& 1.06 & -0.30\\ 
 1849+670& Q &0.657 & -12.83& Mg\,{\sc ii}, H$\beta$ & SK93a & 0.39 & -0.76\\ 
 1928+738& Q &0.302 & -11.29& C\,{\sc iv}, H$\gamma$, H$\beta$, H$\alpha$ &M96,L96 & 3.34 & -0.01\\ 
 1936+714& Q &1.864 & -13.11& C\,{\sc iv} & S93b & 0.34 & -0.35\\ 
 1945+725& Q &0.303 & -12.99& H$\gamma$, H$\beta$, H$\alpha$ & SK93a & 0.28 & -0.81\\ 
 1954$-$388& Q &0.626 & -13.02& H$\beta$ & T93& 2.06 & 0.43\\ 
 1954+513& Q &1.230 & -12.55& Mg\,{\sc ii}, H$\gamma$ &L96 & 1.43 & -0.01\\ 
 1958$-$179& Q &0.650 & -13.09& Mg\,{\sc ii} & O84& 1.20 & 0.13\\ 
 2000$-$330& Q &3.777 & -12.44& Ly$\alpha$, C\,{\sc iv} & O94& 1.15 & 0.78\\ 
 2015+657& Q &2.845 & -13.44& Ly$\alpha$, C\,{\sc iv} & SK93b & 0.51 & -0.70\\ 
 2029+121& BL &1.215 & -14.04& Mg\,{\sc ii} & SK93a& 1.33 & 0.74\\ 
 2044$-$027& Q &0.942 & -12.57& Mg\,{\sc ii} & SS80& 1.02 & -0.52\\ 
 2111+801& Q &0.524 & -13.37& Mg\,{\sc ii}, H$\gamma$, H$\beta$ & SK93a & 0.26 & -0.37\\
 2113+293& Q &1.514 & -13.36& Mg\,{\sc ii} & S93b & 1.47 & 0.62\\ 
 2126$-$158& Q &3.266 & -12.25& Ly$\alpha$, C\,{\sc iv} & O94 & 1.28 & 0.14\\ 
 2128$-$123& Q &0.501 & -11.57& Ly$\alpha$, C\,{\sc iv}, Mg\,{\sc ii},H$\beta$ &T93,O84,O94 & 2.07 & 0.11\\ 
 2134+004& Q &1.936 & -12.13& Ly$\alpha$, C\,{\sc iv}, Mg\,{\sc ii} &B89,O94 &12.30 & 0.83\\ 
 2135$-$147& Q &0.200 & -11.34& Ly$\alpha$, C\,{\sc iv}, H$\gamma$, H$\beta$, H$\alpha$ & JB91,O94,M96& 1.41 & -0.70\\ 
 2136+141& Q &2.427 & -12.65& Ly$\alpha$, C\,{\sc iv} &B81 & 1.11 & -0.10\\ 
 2145+067& Q &0.990 & -11.93& Ly$\alpha$, C\,{\sc iv}, Mg\,{\sc ii} & N79,O94& 4.00 & 0.25\\ 
 2155$-$152& Q &0.672 & -13.59& Mg\,{\sc ii}, H$\beta$ & S89& 1.77 & 0.15\\ 
 2201+315& Q &0.298 & -11.00& Ly$\alpha$, C\,{\sc iv}, Mg\,{\sc ii}, H$\gamma$, H$\beta$, H$\alpha$ &JB91,W95 & 2.32 & 0.24\\ 
 2203$-$188& Q &0.619 & -13.05& Mg\,{\sc ii} & S89& 4.38 & -0.28\\ 
 2209+080& Q &0.484 & -12.53& Mg\,{\sc ii}, H$\beta$ & RS80& 1.08 & -0.39\\ 
 2216$-$038& Q &0.901 & -11.82& Ly$\alpha$, C\,{\sc iv}, Mg\,{\sc ii} &B89,W95 & 1.50 & 0.48\\ 
 2218+395& Q &0.655 & -12.98& Mg\,{\sc ii}, H$\beta$ & SK93b & 0.69 & -0.60\\ 
 2223$-$052& Q &1.404 & -12.46& Ly$\alpha$, C\,{\sc iv}, Mg\,{\sc ii} &P89,W95 & 4.51 & -0.11\\ 
 2223+210& Q &1.949 & -11.92& Ly$\alpha$, C\,{\sc iv} & RS80,W84& 1.24 & -0.65\\ 
 2230+114& Q &1.037 & -11.87& Ly$\alpha$, C\,{\sc iv} & W95& 3.61 & -0.50\\ 
 2234+282& Q &0.795 & -12.95& Mg\,{\sc ii}, H$\gamma$, H$\beta$ & RS80,JB91& 1.06 & 0.08\\ 
 2240$-$260& BL &0.774 & -13.92& Mg\,{\sc ii} & S93a& 1.03 & -0.08\\ 
 2243$-$123& Q &0.630 & -11.94& Mg\,{\sc ii}, H$\beta$ & O84,T93& 2.45 & -0.18\\ 
 2247+140& Q &0.237 & -12.39& H$\gamma$, H$\beta$, H$\alpha$ & GW94& 1.02 & -0.63\\ 
 2251+158& Q &0.859 & -11.88& Ly$\alpha$, C\,{\sc iv}, Mg\,{\sc ii}, H$\gamma$, H$\beta$ & N79,W95,JB91&17.42 & 0.64\\ 
 2255+416& Q &1.149 & -13.81& Mg\,{\sc ii} & SK93a& 1.00 & -0.55\\ 
 2311$-$452& Q &2.884 & -13.79& Ly$\alpha$, C\,{\sc iv} & S89& 1.47 & -0.38\\ 
 2311+469& Q &0.741 & -12.46& Mg\,{\sc ii}, H$\gamma$, H$\beta$ & SK93b & 0.73 & -0.69\\ 
 \end{tabular}
\end{minipage}
\end{table}

\begin{table}
 \begin{minipage}{150mm}
  \contcaption{Data of the sample.}
  \begin{tabular}{ccccllrr}\hline
 2318+049& Q &0.623 & -12.51& Mg\,{\sc ii} & RS80& 1.13 & -0.11\\ 
 2319+272& Q &1.253 & -13.07& Mg\,{\sc ii} & S93b& 1.07 & 0.20\\ 
 2326$-$477& Q &1.306 & -12.14& H$\alpha$ & E89& 2.33 & -0.06\\ 
 2328+107& Q &1.489 & -12.92& C\,{\sc iv} & W84 & 1.01 & -0.05\\ 
 2342+821& Q &0.735 & -13.99& Mg\,{\sc ii}, H$\gamma$, H$\beta$ & L96 & 1.33 & -0.89\\ 
 2344+092& Q &0.673 & -11.80& Ly$\alpha$, C\,{\sc iv}, Mg\,{\sc ii}, H$\gamma$, H$\beta$ & B89,O94,JB91 & 1.43 & -0.09\\ 
 2345$-$167& Q &0.576 & -12.75& H$\gamma$, H$\beta$ & JB91 & 3.66 & 0.51\\ 
 2351+456& Q &1.992 & -13.58& C\,{\sc iv}, Mg\,{\sc ii} & L96 & 1.42 & -0.05\\ 
\hline
\end{tabular}
\end{minipage}

\medskip

Notes for the table 1. Q: quasars; BL: BL Lac objects.\\
Column (1): IAU source name. Column (2): classification of the source.
Column (3): redshift. Column (4): estimated total
broad-line flux (erg s$^{-1}$ cm$^{-2}$). Column (5): lines from which 
the total $f_{line}$ has been estimated. Column (6): references for the line 
fluxes. Column (7): radio flux density at 5 GHz. Column (8): two-point 
spectral index between 6$-$11 cm.\\

\vskip 1mm
References:
B81: Baldwin et al. (1981). 
B89: Baldwin et al. (1989).
B94: Baker et al. (1994). 
BK84: Bergeron \& Kunth (1984). 
C91: Corbin (1991).
CF94: Corbin \& Francis (1994). 
d94: di Serego Alighieri et al. (1994). 
E89: Espey et al. (1989). 
F83: Fricke et al. (1983). 
GW94: Gelderman \& Whittle (1994). 
H78: Hunstead et al. (1978). 
JB91: Jackson \& Browne (1991). 
K85: Kinney et al. (1985). 
L96: Lawrence et al. (1996). 
M96: Marziani et al. (1996). 
N79: Neugebauer et al. (1979). 
O94: Osmer et al. (1994). 
O84: Oke et al. (1984).
P89: Perez et al. (1989).
R84: Rudy (1984). 
RS80: Richstone \& Schmidt 1980. 
S81: Steiner (1981).
S89: Stickel et al. (1989).
S93a: Stickel et al. (1993a). 
S93b: Stickel et al. (1993b). 
SJ85: Stiko \& Junkkarinen (1985). 
SK93a: Stickel \& K\"uhr (1993a). 
SK93b: Stickel \& K\"uhr (1993b). 
SM87: Stockton \& MacKenty (1987). 
SS80: Smith \& Spinrad (1980). 
T93: Tadhunter et al. (1993). 
V95: Vermeulen et al. (1995). 
W84: Wampler et al. (1984).  
W85: Wills et al. (1985). 
W95: Wills et al. (1995). 
W98: Wang et al. (1998).
WM96: Warren \& Moller (1996).   
\end{table}

\section{Results}

We present the radio luminosity at 5 GHz and the total
broad-line luminosity as functions of redshift $z$ for our sample
in Figs. 1 and 2 respectively. The radio luminosity has been K-corrected, 
and the low flux limit is indicated by the dashed line in Fig. 1 assuming the
spectral index $\alpha=$ 0. We note that our sample provide  wide
dispersion in both the radio and broad-line luminosities at
any redshift $z>$0.3.

 Figure 3 shows the correlation between radio flux $\nu f_{\nu 5G}$ at
 5 GHz (K-corrected) and the total broad-line flux $f_{\rm line}$. 
We test the radio flux and the total broad-line flux against each other
and find that the distribution of the radio flux is significantly different
from that of the total broad-line flux (99.98 per cent confidence by
Kolmogorov-Smirnov test). 
We find the radio and the total broad-line fluxes are significantly
correlated for quasars in the sample ($\gg$99.99 per cent confidence
using Spearman's correlation coefficient $\rho$). The result is same
for the whole sample (we exclude the source 1226+023 in the analyses).
The BL Lac objects seem to follow the similar
statistical behaviour of the quasars, but with relatively lower 
broad-line fluxes (see Fig. 3). Fig. 4 shows strong correlation between
the radio and broad-line luminosities. 
We also performed analyses on a subsample in the restricted redshift range: 
$0.5<~z<1.5$. A correlation between the radio and broad-line fluxes  
is found at 99.93 per cent, while the correlation between luminosities 
is at 99.7 per cent. 

There are 32 steep-spectrum quasars with  radio spectral index
$\alpha_{11-6}\le -0.7$ in our sample(marked by squares in figures).
The sample without these steep-spectrum quasars is similar to those 
for the whole sample.  No significant correlation
is found for this steep-spectrum quasar subsample between fluxes, while a
weak correlation is found between luminosities. However, we have not found
obvious different behaviours between the steep-spectrum quasars and the
remains of the sample, though detailed comparisons are not possible
due to the small number of steep-spectrum quasars.

\section{Discussion}
The radio emission for steep-spectrum quasars is believed to
be unbeamed emission from the lobes. 
The dominant influence on the radio luminosity is $Q_{jet}$, and not the
large-scale radio source environment (Serjeant et al. 1998), though
$Q_{jet}$ also depends on the gaseous environment of the radio source
(Ellingson et al. 1991; Rawlings \& Saunders 1991).
Therefore the radio emission can be a measure of the bulk power in the jets
$Q_{jet}$ (Serjeant et al. 1998).
The optical continuum  is a good indicator of the disc surrounding
a black hole for the steep-spectrum quasars, 
since relativistic beaming does not affect the optical
continuum in these sources.
Thus, the radio-optical correlation gives evidence of
a disc-jet link (Serjeant et al. 1998).

The optical radiation may
be enhanced by relativistically beamed synchrotron radiation for some
flat-spectrum quasars, which prevent us from using the optical continuum 
as an indicator of the disc. The broad-line region is photoionized by
a nuclear source (probably radiation from the disc), so
the broad-line emission instead of the optical continuum can be used as
an indictor of the accretion power for both steep and flat-spectrum quasars
(Celotti, Padovani \& Ghisellini 1997).
The $Q_{jet}$ is proportional to the bulk Lorentz factor of the jet,
and can be estimated if the size of the jet and density of the
electrons in the jet are available(Celotti \& Fabian 1993).
The radio emission for flat-spectrum quasars is beamed 
and  is thought to be mainly from the core of the jet. The Doppler factor
of the jet is the dominant influence on the observed radio luminosity.
The radio luminosity is also determined by the size of the
jet,  the electron density, and the magnetic energy density in the jet.
The Doppler factor depends on both the values of Lorentz factor and 
viewing angle of the jet. The viewing angle of the jet for flat-spectrum
quasar is believed to be smaller that that of steep-spectrum quasar.
The radio luminosity of flat-spectrum quasars can be naturally linked
to the jet power, though some uncertainties are induced mainly by the
unknown viewing angle.
Part of the scatter in the radio and broad-line emission correlations
can be attributed to the uncertainty of viewing angle for flat-spectrum
quasars. 
The radio luminosity seems to be a good indicator
for the jet in both steep or flat-spectrum quasars, but in different ways. 
The  correlation found in this work between radio and broad-line emission
suggests a close link between formation of the jets and accretion on
to the central black hole. Another possible interpretation of the
correlation is that the broad-line region is photoionized by the
relativistic beamed radiation from the jets. However, the similarity of
the broad-line component in radio-quiet and radio-loud sources and
no systematically larger equivalent widths for lines in steep-spectrum
quasars than that in  flat-spectrum quasars are found (Celotti, Padovani \&
Ghisellini 1997; Ghisellini 1998), which seems to rule out this
possibility. The present radio and broad-line correlation gives a new
evidence for  a close link between the jets and discs, though how the
disc and jet is coupled is still not clear.

\section*{Acknowledgments}
The support from NSFC and Pandeng Project is gratefully acknowledge. 
We thank Peter Scheuer for his helpful suggestions and the 
corrections on the manuscript.
The anonymous referee is thanked for the useful suggestions
that improved the presentation of this paper, and for his prompt reply.
This research has made use
of the NASA/IPAC Extragalactic Database (NED), which is operated by the Jet
Propulsion Laboratory, California Institute of Technology, under contract
with the National Aeronautic and Space Administration.

{}

\end{document}